\documentclass[12pt,letterpaper]{article}
\usepackage{graphicx}
\usepackage{dcolumn}
\usepackage{bm}
\usepackage{amssymb}
\usepackage{amsmath}
\usepackage{amsfonts}
\usepackage{setspace}
\usepackage[left=1.25in, right=1.25 in]{geometry}

\def\dsum{\sum}

\begin{document}

\title{Partial Kekule Ordering of Adatoms on Graphene}

\author{V.~V.~Cheianov$^1$, V.~I.~Fal'ko$^1$, O.~Sylju\aa sen$^2,$ and B.~L.~Altshuler$^{3}$ }
\maketitle

\footnotetext[1]{Physics Department, Lancaster University, Lancaster
LA1 4YB, UK }

\footnotetext[2]{Department of Physics, University of Oslo, PO Box
1048 Blindern, N-0316 Oslo, Norway}

\footnotetext[3]{Physics Department, Columbia
University, 538 West 120th Street, New York, NY 10027, USA}

\begin{abstract}
Electronic and transport
properties of Graphene, a one-atom thick crystalline material,
are sensitive to the presence of atoms adsorbed on its surface. An ensemble
of randomly positioned adatoms, each serving as a scattering center,
leads to the Bolzmann-Drude diffusion of charge determining the resistivity of
the material.
An important question, however, is whether the distribution of adatoms is
always genuinely random.
In this Article we demonstrate that a dilute adatoms
on graphene may have a tendency towards a spatially
correlated state with a
hidden Kekul\'e mosaic order. This effect emerges from the interaction
between the adatoms mediated by the Friedel oscillations of the
electron density in graphene.
The onset of the ordered state, as the system is cooled below the critical
temperature, is accompanied by the opening of
a gap in the electronic spectrum of the material, dramatically changing its transport
properties.
\end{abstract}

\bigskip

When can an apparently random system be considered ordered? Or can an apparently random ensemble of
impurities in a system be correlated enough to force the reconstruction of the electronic band structure in a
material? In this Article we predict that a dilute ensemble of adatoms sprinkled randomly over a graphene
monolayer \cite{Geim1,GraFET} can establish long-range correlations between their
positions, despite the fact that they may be many graphene unit-cell lengths apart. This correlation is
strong enough that at a transition temperature it will induce an energy gap in the electronic spectrum
despite the fact that, in the ``ordered'' state, the distribution of adatoms does not show any crystalline
structure. It rather resembles the ferromagnetically ordered state of spins of magnetic ions in dilute
magnetic semiconductors \cite{GaMnAs}. The physical mechanism behind this phenomenon is the electron-mediated
interaction between the adsorbents, which prompts their partial ordering into a configurations associated
with a hexagonal superlattice with a unit cell three times bigger than that of graphene. Since the density,
$\rho $ of adsorbents is low, they occupy a small randomly chosen fraction
of the equivalent positions on the
superlattice. This ordering folds the Brillouin zone and thus opens a
spectral gap for low-energy electrons.
This phenomenon suggests a novel route towards engineering the band structure and controlling transport in
graphene-based devices.

Graphene \cite{Dresselhaus,GraFET} is a two dimensional crystal of
carbon atoms, which form a honeycomb lattice with two distinct
sublattices (A and B). The first Brillouin zone (BZ) has a hexagonal
form (the blue area in Fig.
1A), and the conduction band touches the valence band in six BZ corners \cite%
{Dresselhaus} which form two non-equivalent triads of BZ corners, $%
\mathbf{K}$ and $\mathbf{K}^{\prime }$ connected by the reciprocal lattice
vectors, $\mathbf{G}$ and $\mathbf{G}^{\prime }$. Low-energy electronic
excitations in the momentum space are located in the vicinities of the points
$\mathbf{K}$ and $\mathbf{K}^{\prime }$, i.e. belong to one of the two
valleys with the linear 'Dirac-type' spectrum, $\varepsilon (p)=\pm vp$
where $\mathbf{p}$ is the momentum counted from one of the $\mathrm{K}$%
-points and $v\approx 10^{8}$cm/sec. Both the gapless spectrum and the
valley degeneracy follow directly from the symmetries of the honeycomb
lattice. In pristine graphene, the honeycomb lattice is stable against
spontaneous structural changes.

Recently, several types of adatoms were used to dope graphene in attempts
to tailor properties of graphene-based devices \cite%
{Geim2,Fuhrer,Lanzara,Geim3}. Below, we consider theoretically a
particular example, Fig. 2A, of adatoms, such as alkali atoms
\cite{K,Li}, Ca or Al whose stable positions are above the centers of the
hexagons. A single adatom of this type preserves rotational and
reflection symmetries but breaks the translational symmetry of the
lattice. Therefore, it can scatter electrons between valleys. The
intervalley scattering generates the Friedel oscillations (FO) of
the electron density, which amplitude rather slowly decays when the
distance from the adatom \cite{FO}. The pattern and period of such
FO are
determined by the wave vectors $\delta \mathbf{K}=\pm (\mathbf{K}^{\prime }-%
\mathbf{K)}$ transferred upon scattering between the valleys. Due to a
peculiar relation of $\mathbf{K}$, $\mathbf{K}^{\prime }$ and the reciprocal
lattice vectors in graphene, $\mathbf{K}=\frac{1}{3}(\mathbf{G}+\mathbf{G}^
\prime )$ and $\mathbf{K}^{\prime
}=-\frac{1}{3}(\mathbf{G}+\mathbf{G}^\prime )$, the FO formed around
an adatom have a structure of a charge density wave with the
hexagonal lattice pattern and the unit cell extended over three unit
cells of graphene. In graphene with zero carrier density ($\rho
_{e}=0$), the amplitude of these superlattice oscillations decays as
inverse cube of the distance between the adatoms. The oscillations
of the electron density induced by one adatom affect other add
atoms, thus leading to the pairwise interaction between adsorbents,
which is sensitive to their position in the superlattice. In Fig. 2
(A and B), we compare the potential landscape for a probe adatom
created by several others due to their FO, to stress that its
amplitude is substantially enhanced by ordering.
Note that the long-range RKKY-type \cite{RKKY} interaction in a
low-density 'gas' of adatoms as well as the ordering it promotes
has little to do with those in dense aggregates of adsorbents.
Indeed, the interaction at atomic distances is mediated by local
lattice deformations - phonons. Such an interaction decays
exponentially as a function of distance between the adatoms, as
opposed to the power law $1/R^3$ decay of the electron-mediated
coupling.

 Figure 2 illustrates an example of hidden structural
ordering of adatoms sprinkled over graphene. Without mosaic
coloring (or a superlattice mesh) it would be difficult to
distinguish the ordered configuration of adatoms (Fig. 2D) from a
disordered state (Fig. 2C). With the help of colors, one can
identify a\ triple-size unit cell of the superlattice, with three
non-equivalent adatom positions (red, blue, and green) between six
carbons, which resembles a hexagonal Kekul\'{e}-type \cite{Kekule}
lattice \cite{Kekule1,Li}. The FO-mediated interaction $V_{ij}$
between two adatoms $i$ and $j$ depends on whether they occupy
equivalent (same color) positions on the superlattice, or not. This
consideration maps the problem of hidden Kekul\'{e} mosaic ordering
of adsorbents onto the three-state Potts model \cite{Wu} with a
random in strength 'ferromagnetic' coupling of species.

To estimate critical temperature of the hidden ordering, one has to
evaluate the function $V_{ij}(\mathbf{r}_{i}-\mathbf{r}_{j})$, where
$\mathbf{r}_{i}$ and $\mathbf{r}_{j}$ are adatoms locations. We
use the technique developed for the studies dealing with disorder in
graphene \cite{Impurities,DisorderedG,FO} and describe electrons by
four-component Bloch function $\psi $ (taking into account its
sublattice composition and valley degeneracy) and the 4x4
Hamiltonian,
\begin{equation}
H=v(\mathbf{p}\cdot \mathbf{\sigma })\Pi _{z}+\lambda hva\dsum\limits_{i}%
\mathbf{(\mathbf{u}_{i}\cdot \Pi })\delta (\mathbf{r}-\mathbf{r}_{i}).
\label{H}
\end{equation}%
Here, Pauli matrices $\sigma _{x,y,z}$ act on sublattice indices and $\Pi
_{x,y,z}$ on the valley indices of $\psi $ \cite{Impurities,DisorderedG,FO}.
The first term in Eq. (\ref{H}) is a familiar Hamiltonian for Dirac-type
electrons in pristine graphene. The second accounts for the intervalley
scattering of electrons by adatoms, with $\lambda $ being the
dimensionless coupling constant (realistically, $\lambda \lesssim 1$ since $%
hv/a$ is of the order of the bandwidth in graphene). Unit two-component
vector $\mathbf{\mathbf{u}}_{i}=(\cos \frac{2\pi m_{i}}{3},\sin \frac{2\pi
m_{i}}{3})$ specifies which of the three non-equivalent positions the $i$-th
adatom occupies on the superlattice, with \ $m_{i}=-1$, $0$ and $-1$ (red,
blue, and green hexagons).

Using thermodynamic perturbation theory and the standard RKKY approach \cite%
{RKKY} we express the interaction between adsorbents mediated by electrons
as
\begin{eqnarray}
V_{ij} &=&2(\lambda hva)^{2}\mathrm{Tr}\int_{-\infty }^{\infty }d\tau
\mathbf{(\mathbf{u}_{i}\cdot \Pi })G(\mathbf{r}_{i}-\mathbf{r}_{j},\tau )%
\mathbf{(\mathbf{u}_{i}\cdot \Pi })G(\mathbf{r}_{j}-\mathbf{r}_{i},-\tau ),
\nonumber \\
G(\mathbf{r},\tau ) &=&-\frac{1}{4\pi }\frac{v\tau +i\Pi _{z}\mathbf{\sigma }%
\cdot \mathbf{r}}{(v^{2}\tau ^{2}+r^{2})^{3/2}},
\end{eqnarray}%
where $\tau$ is imaginary time and $G(\mathbf{r},\tau )$ is the
zero-temperature Green function of Dirac-like electrons. Strictly
speaking the equation 2 is valid at $T=0.$ However one can use it as
long as  $ T<hv\rho ^{1/2}.$ The, trace ($\mathrm{Tr}$) is taken
over the sublattice and valley indices. The electron spin degeneracy
is accounted for by the overall factor of 2. The resulting
electron-mediated 'ferromagnetic' interaction between adatoms at a
distance $|\mathbf{r}_{i}-\mathbf{r}_{j}|\gg a$,
\begin{eqnarray}
V_{ij} &=&-J\frac{\mathbf{u}_{i}\cdot \mathbf{u}_{j}}{|\mathbf{r}_{i}-%
\mathbf{r}_{j}|^{3}\rho ^{3/2}},\;\;\mathbf{u}_{i}\cdot \mathbf{u}_{j}=\cos
\frac{2\pi (m_{i}-m_{j})}{3},  \label{Vij} \\
J &=&\frac{\lambda ^{2}}{2}(a^{2}\rho )^{3/2}\frac{hv}{a},  \nonumber
\end{eqnarray}%
has a long-range tail, $V\propto |\mathbf{r}_{i}-\mathbf{r}_{j}|^{-3}$. The
typical interaction energy scale, $J$ is the interaction strength at a mean
distance between the nearest neighbors, $\sim \rho ^{-1/2}$ (recall that $%
\rho \ll a^{-2}$).

To evaluate the critical temperature, $T_{c}$, we modeled the ordering
transition numerically. We used the cluster Monte Carlo algorithm \cite%
{ClusterMonteCarlo} to compute statistical moments
\[
M_{2n}=\frac{\sum_{\mathbf{u}_{1},...,\mathbf{u}_{N}}\left( \sum_{i=1}^{N}%
\frac{\mathbf{u}_{i}}{N}\right) ^{2n}e^{-\frac{1}{2T}\sum V_{ij}}}{\sum_{%
\mathbf{u}_{1},...,\mathbf{u}_{N}}e^{-\frac{1}{2T}\sum V_{ij}}},
\]%
for 10 realizations of quenched Poissonic distributions of
$N=2\times 10^{4}$ adatoms. The ordering transition can be
detected by a sudden rise of the order parameter $M\equiv
\sqrt{M_{2}}$, from $M(T>T_{c})=0$ to $M(T<T_{c})=1$ accompanied by
decrease of $\eta =M_{4}/M_{2}^{2}$, from $\eta (T\gg T_{c})=2 $
(set by the central limit theorem for a large number of uncorrelated
clusters) to $\eta (T\ll T_{c})=1$. Results of the numerical
analysis are presented in Fig. 3. The transition temperature turned
out to be
\begin{equation}
T_{c}\approx 8J\sim 4\lambda ^{2}(a^{2}\rho )^{3/2}\frac{hv}{a}.  \label{Tc}
\end{equation}%
For example, for $\lambda \sim 1$, just 1\% coverage of graphene by
adatoms should generate $T_{c}$ in the room temperature range.

Since the mobility of adatoms on graphene strongly depends on
temperature, the higher the adsorbent density $\rho $, the higher
$T_{c}$ is, and the quicker the self-organization should establish
upon cooling. Note that the aggregation of adsorbents, such as
discussed in Refs. \cite{Aggregation, Levitov}, would be a much
slower process
in a dilute system.  At $T<T_{c}$ the proposed partial ordering
suppresses adatoms diffusion leading to a further slowdown of
aggregation.

The value of $T_{c}$ may also depend on the concentration $\rho
_{e}$ of electrons (or holes) in graphene. Finite carrier density
leads to the additional modulation of the FO, with the period twice
as small as the electron Fermi wavelength $\sim\rho_e^{-1/2}$
\cite{FO}. For $\rho _{e}\gtrsim \rho $ these modulations would make
the sign of the interaction between adatoms random and, thus
eliminate the ordering. Therefore it seems to be possible to control
hidden ordering of adsorbents electrically, by filling or depleting
the flake with carriers - the method already in use to fine-tune the
ferromagnetic transition temperature in thin films of dilute
magnetic semiconductors \cite{OhnoDietl}.

The self-organization of an apparently random ensemble of adatoms into a
Kekul\'{e}-type ordered state drastically changes electronic spectrum in graphene.
Adatoms that preferentially occupy one of the three
equivalent positions in the supercells over a length scale $L\gg \sqrt{%
1/\rho }$ can Bragg scatter electrons between the two valleys coherently.
This implies the Brillouin zone folding in Fig. 1 (B and C): all of the
points $\mathbf{K}$ and $\mathbf{K}^{\prime }$ of the original BZ are
projected onto the $\Gamma $-point of a smaller BZ corresponding to the
superlattice with a triple unit cell. Simultaneously, a gap, $\Delta, $ opens
in the electronic spectrum
\begin{equation}
\varepsilon (p)=\pm \sqrt{(vp)^{2}+\Delta ^{2}},\;\Delta =\lambda a^{2}\rho
\frac{hv}{a}.  \label{gapped}
\end{equation}%
To derive Eq. (\ref{gapped}), one can substitute the second term of the
Hamiltonian in Eq. (\ref{H}) by its average, for example, for all adatoms
positioned on yellow hexagons, and diagonalize the resulting matrix, $\bar{H}%
=v\mathbf{\sigma }\cdot \mathbf{p}+\lambda va\rho \Pi _{x}$.

One can think of several ways to experimentally detect the hidden Kekul\'{e}
mosaic order. One is to use the angle-resolved photoemission spectroscopy
(ARPES). The latter technique is not only a natural method to
reveal the formation of the spectral gap. It can also provide a direct
evidence of the BZ folding. Indeed, ARPES measures simultaneously the energy
and all three momentum components of the photo-emitted electrons. While at
low energies only the vicinity of the BZ corners $\mathbf{K}$ and $\mathbf{K}%
^{\prime }$ can be seen in pristine graphene \cite{ARPES,Mucha}, the Bragg
scattering by the self-organized adsorbents generates an ARPES signal also
at the $\Gamma $-point of the BZ in Fig. 1A. Another signature of the Kekul%
\'{e} ordering would be a bright appearance of the D peak in the
phonon Raman scattering \cite{Raman}: the excitation of a BZ phonon
forbidden by momentum conservation in pristine graphene. Finally,
the gap in the electronic spectrum would dramatically affect charge
transport in graphene. This may offer numerous opportunities for
graphene-based electronics.

\section*{Acknowledgements}
The authors would like to thank George Pickett for
valuable comments. The work was supported by the Lancaster-EPSRC Portfolio
Partnership, ESF CRP SpiCo, and
US DOE contract No. DE-AC02-06CH11357. Numerical computations were
carried out using resources provided by the Notur project of the
Norwegian Research Council.

\pagebreak

\section*{Figure legends}
\vskip 20 pt

\noindent Figure 1: (A) Brillouin zone of graphene in the reciprocal space is shown as a
blue hexagon. The
two valleys $K$ and $K'$ are situated in the six corners of the hexagon, which
are identified via reciprocal lattice translations generated by vectors $\mathbf G$ and
$\mathbf G'.$
(B) Brillouin zone folding due to the ordering transition.
The folding leads to the identification of the valley points $K$ and $K'$ with the
$\Gamma$ point in the center of the Brillouin zone.
(C) The energy surface in the folded Brillouin zone. Due to the
interaction between the valleys a gap opens in the spectrum.

\vskip 80 pt

\noindent Figure 2: Kekul\'e mosaic ordering of adatoms of  {\it the same chemical
element} on graphene lattice. Panels (A) and (B) show the potential
landscape that an extra atom would see in the presence of four atoms
already shown. Coloring of the atoms is introduced to reveal their
position within the Kekul\'e superlattice, as shown in panels (C) and
(D). From a comparison of (A) and (B) one can see that
adatoms placed on unicolor tiles enhance the potential landscape
forcing other atoms to occupy tiles of the same color.

\pagebreak

\noindent Figure 3: The
Kekul\'e mosaic order parameter $M$ as a function of temperature.
The phase transition to the ordered state is characterized by a rise
of $M$ accompanied by a sharp drop of the measure of finite-size
fluctuations $M_4/M_2^2.$ The data were obtained using the cluster
Monte Carlo algorithm. The error bars indicate the standard
deviation of the thermodynamic quantities in an ensemble of 10
random realizations of Poisson distributions of $N=2\times 10^4$
atoms in the plane.

\end{document}